\documentclass[11pt]{article}
\usepackage{a4wide}                      % to get a wider page margen
\usepackage{epsf}                        % to make figures
\usepackage{latexsym}                    % to make fancy symbols
\usepackage{amsfonts}                    % to make fancy symbols
\usepackage{amssymb}                     % to make fancy symbols
\newcommand{\sect}[1]{\setcounter{equation}{0}\section{#1}}

\def\be{\begin{equation}}
\def\ee{\end{equation}}
\def\bea{\begin{eqnarray}}
\def\eea{\end{eqnarray}}

\def\part{\partial}

\def\incl{\mbox{i}}

\def\R{\ensuremath{\mathbb{R}}}
\def\C{\ensuremath{\mathbb{C}}}

\def\makeatletter{\catcode`\@=11}% 11:letter
\makeatletter
\def\mathbox#1{\hbox{$\m@th#1$}}%
\def\math@ccstyles#1#2#3#4#5#6#7{{\leavevmode
      \setbox0\mathbox{#6#7}%
      \setbox2\mathbox{#4#5}%
      \dimen@ #3%
      \baselineskip\z@\lineskiplimit#1\lineskip\z@
      \vbox{\ialign{##\crcr
             \hfil \kern #2\box2 \hfil\crcr
             \noalign{\kern\dimen@}%
             \hfil\box0\hfil\crcr}}}}
\def\mathaccstyles{\math@ccstyles\maxdimen}
\def\maththroughstyles{\math@ccstyles{-\maxdimen}}
\def\unity%
 {\maththroughstyles{.45\ht0}\z@\displaystyle {\mathchar"006C}\displaystyle 1}
\begin{document}

\rightline{UG-FT-175/04}
\rightline{CAFPE-45/04}
\rightline{FFUOV-04/17}
\rightline{QMUL-PH-04-10}
\rightline{hep-th/0411181}
\rightline{January 2005}
%\vspace{2truecm}
\vspace{1truecm}

%%%%%%%%%%%%%%%%%
\centerline{\LARGE \bf Giant Gravitons and Fuzzy $CP^2$}
\vspace{1.3truecm}

\centerline{
    {\large \bf Bert Janssen${}^{a,}$}\footnote{E-mail address:
                                  {\tt bjanssen@ugr.es}},
    {\large \bf Yolanda Lozano${}^{b,}$}\footnote{E-mail address:
                                  {\tt yolanda@string1.ciencias.uniovi.es}}
    {\bf and}
    {\large \bf Diego Rodr\'{\i}guez-G\'omez${}^{b,c,}$}\footnote{E-mail address:
                                  {\tt diego@fisi35.ciencias.uniovi.es}}
                                                            }

\vspace{.4cm}
\centerline{{\it ${}^a$ Departamento de F\'{\i}sica Te\'orica y del Cosmos and}}
\centerline{{\it Centro Andaluz de F\'{\i}sica de Part\'{\i}culas Elementales}}
\centerline{{\it Universidad de Granada, 18071 Granada, Spain}}

\vspace{.4cm}
\centerline{{\it ${}^b$Departamento de F{\'\i}sica,  Universidad de Oviedo,}}
\centerline{{\it Avda.~Calvo Sotelo 18, 33007 Oviedo, Spain}}

\vspace{.4cm}
\centerline{{\it ${}^c$Queen Mary, University of London,}}
\centerline{{\it Mile End Road, London E1 4NS, UK}}

\vspace{2truecm}

%%%%%%%%%%%%%%%%%
\centerline{\bf ABSTRACT}
\vspace{.5truecm}

\noindent In this article we describe the giant graviton configurations in $AdS_m\times S^n$
backgrounds
 that involve 5-spheres,
namely, the giant graviton in $AdS_4\times S^7$ and the dual giant graviton in
$AdS_7\times S^4$, in terms of dielectric gravitational waves. Thus, we conclude
the programme initiated in hep-th/0207199 and pursued in hep-th/0303183 and
hep-th/0406148 towards the microscopical description of giant gravitons in 
$AdS_m\times S^n$ spacetimes.
In our construction the
gravitational waves expand due to Myers dielectric  effect onto
``fuzzy 5-spheres'' which
are described as $S^1$ bundles over fuzzy $CP^2$. These fuzzy manifolds
appear as solutions of the matrix model that comes up as the action for M-theory
gravitational waves.
The validity of our
description is checked by confirming the agreement with the Abelian
description in terms of a spherical M5-brane
when the number of waves goes to
infinity.

%%%%%%%%%%%%%%%%%%%%%%%%%%%%%%%%%%%%%%%%%%%%%%%%%%%%%%%%%%%%%%%%%%%%%%%%%%%%%%%%%%%%%

\newpage
\section{Introduction}

It is by now well-known that giant gravitons can be described microscopically in terms
of dielectric \cite{myers}
gravitational waves \cite{DTV,BMN,JL2,JLR,JLR2}.  In 
$AdS_m\times S^n$
spacetimes the gravitational waves expand into a fuzzy $S^{(n-2)}$ brane included in
$S^n$, for the genuine giant graviton, 
 or into a fuzzy $S^{(m-2)}$ brane included in $AdS_m$, for the dual giant graviton, both 
 carrying angular momentum on the spherical part of the geometry. 
 For the giant graviton in 
 $AdS_7\times S^4$ and the dual giant graviton in $AdS_4\times S^7$ these fuzzy spheres
 are ordinary non-commutative $S^2$ \cite{JL2}, whereas for both the giant and dual giant
 gravitons in $AdS_5\times S^5$ the corresponding ``fuzzy 3-spheres'' are defined as
 $S^1$ bundles over non-commutative $S^2$ base manifolds \cite{JLR}.
 In all cases
 perfect agreement is found, when the number of waves goes to infinity,
 with the (Abelian) description in \cite{GST,GMT,HHI}  in
 terms of spherical test branes.
 This agreement provides the strongest support to the
 (non-Abelian) dielectric constructions of \cite{JL2,JLR}. 
 
 The key point in the non-Abelian description of giant gravitons is the construction of the action for the
 system of coincident waves, and the identification of the dielectric and magnetic moment
 couplings responsible of their expansion. Given that 
 $AdS_m\times S^n$ is not weakly curved one cannot use linearised  Matrix
 theory results as those in \cite{TVR1,TVR2}.
 
 The action appropriate to describe  coincident gravitational waves in non-weakly
 curved M-theory backgrounds was constructed in \cite{JL2}. This action contains
 dielectric and magnetic moment couplings to the 3-form potential of eleven dimensional
 supergravity, which are responsible for the expansion of the waves into a dielectric
 or magnetic moment M2-brane with the topology of a fuzzy $S^2$, which constitute,
 respectively,  the
 dual giant and giant graviton configurations in the $AdS_4\times S^7$ and $AdS_7\times S^4$
 backgrounds. Using this action one can also describe coincident Type IIB gravitational
 waves in non-weakly curved $AdS$ type backgrounds, after reduction and T-duality \cite{JLR}.
 In the Type IIB action the T-duality direction occurs as an special isometric direction, 
 and this turns out to
 be essential in the construction of  the fuzzy manifold associated to both giant and dual
 giant graviton configurations in $AdS_5\times S^5$. Describing the ``fuzzy 3-sphere'' as
 an $S^1$ bundle over 
 a fuzzy 2-sphere,  the isometric direction is precisely the coordinate along the fibre.
 This same action was
 later used in \cite{JLR2} to describe the giant graviton solutions in the 
 $AdS_3\times S^3\times T^4$ Type IIB background
 in terms of expanding waves. In this case
 the waves expand into a fuzzy cylinder whose basis is contained either in $S^3$ or
 in $AdS_3$. In all these cases the agreement between these descriptions and the Abelian
 descriptions of \cite{GST,GMT,HHI} for large number of waves provides the strongest
 support for the validity of the non-Abelian actions for coincident 
 waves\footnote{Of course, together with the 
 fact that the action for
 M-theory waves reduces to Myers action for D0-branes when they propagate along the eleventh
 direction. This was in fact the key ingredient in \cite{JL2} in order to extend the
 linearised Matrix theory result to more general backgrounds. See this reference
 for more details.}.
 
 In this paper we would like to complete the programme initiated in \cite{JL2}, and
 pursued in \cite{JLR} and \cite{JLR2}, with the microscopical description of the giant
 graviton in $AdS_4\times S^7$ and the dual giant graviton in $AdS_7\times S^4$.
 One expects that microscopically these gravitons will be described in terms of
 a magnetic moment or dielectric M5-brane with the topology of a fuzzy 5-sphere.
 Fuzzy $S^n$ with $n>2$ are however quite complicated technically. The general
strategy is to identify them as subspaces of suitable spaces (spaces which admit a
symplectic structure) and to introduce conditions to restrict the functions to be
on the sphere \cite{Ram1,Ram2} \cite{S-J}. 
 We will see that in our construction the ``fuzzy $S^5$'' is simply defined as an
 $S^1$ bundle over a fuzzy $CP^2$, in very much the same way the ``fuzzy $S^3$'' in 
 \cite{JLR} was defined as an $S^1$ bundle over a fuzzy $S^2$. Again the
 coordinate along the $S^1$ fibre is, as in the fuzzy $S^3$ of \cite{JLR}, an
 isometric direction present in the action describing the waves. The existence of this
 direction is, moreover, crucial in order to construct the right dielectric
 couplings that will cause the expansion of the waves. As we will discuss 
the expanded 5-brane will be a longitudinal brane wrapped on this direction.
 
The fuzzy $CP^2$ has been extensively studied in the literature
(see for instance \cite{NR}-\cite{KNR}).
 In the context of Myers dielectric effect it was studied in \cite{TV} \cite{ABS}. 
$CP^2$ is the coset manifold $SU(3)/U(2)$. $G/H$ coset manifolds 
 can be described as fuzzy surfaces if $H$ is the isotropy group of the lowest weight state 
of a given irreducible representation of $G$ \cite{Madore,TV}.
When $G=SU(2)$ all lowest weight vectors have isotropy group $U(1)$, and therefore one describes 
a fuzzy $SU(2)/U(1)$, i.e. a fuzzy $S^2$ for any choice of irreducible representation.
In the case of $SU(3)$ irreducible representations can be parametrised by two integers 
$(n,m)$, corresponding to the number of fundamental and anti-fundamental indices. 
The lowest weight vector in the representations $(n,0)$ or $(0,n)$ has isotropy group $U(2)$, 
whereas for any other irreducible representation it has isotropy group $U(1)\times U(1)$. 
Therefore, choosing a $(n,0)$ or a $(0,n)$ irreducible representation one describes a 
fuzzy $CP^2$.

 We will use this result to describe ``the fuzzy $S^5$'', associated to the giant graviton in 
$AdS_4\times S^7$ and the dual giant graviton in $AdS_7\times S^4$, as an $S^1$ bundle over 
a fuzzy $CP^2$ base manifold. We will then use the action for coincident M-theory 
gravitational waves to find the corresponding ground state configuration.
 A key ingredient in this construction is the identification in the action of the 
direction along the $S^1$ bundle. For this purpose we will start in Section 2 by 
recalling some properties of the action for coincident M-theory gravitational waves 
constructed in \cite{JL2}. Then in Section 3 we will use this action to describe 
microscopically the giant graviton in $AdS_4\times S^7$. We will see that the
corresponding macroscopical description is in terms of a longitudinal M5-brane with
$S^5$ topology.
In Section 4 we present the 
analogous description for the dual giant graviton in $AdS_7\times S^4$. In both cases we 
show the explicit agreement with the macroscopical description in \cite{GST,GMT} for 
large number of gravitons. In Section 5 we present our Conclusions, where we discuss the
connection between our solution and other 5-brane solutions to Matrix
theory actions previously found in the literature, as well as the supersymmetry properties
of our configurations.

 \sect{The action for M-theory gravitational waves}
 
 The action for coincident
 M-theory gravitational waves constructed in \cite{JL2} is given by:
 \begin{equation}
 \label{Mwaves}
 S=S^{\rm BI}+S^{\rm CS}
 \end{equation}
 with BI action given by
 \begin{equation}
 S^{\rm BI}=-T_0 \int d\tau {\rm STr} \{ k^{-1}\sqrt{-P[E_{00}+E_{0i}
 (Q^{-1}-\delta)^i_k E^{kj}E_{j0}]{\rm det Q}}\}\, ,
 \end{equation}
 where
 \begin{equation}
 \label{mred}
 E_{\mu\nu}={\cal G}_{\mu\nu}+k^{-1}(\incl_k C^{(3)})_{\mu\nu}\, , \qquad
 {\cal G}_{\mu\nu}=g_{\mu\nu}-\frac{k_\mu k_\nu}{k^2}
 \end{equation}
 and
 \begin{equation}
 Q^i_j=\delta^i_j+ik[X^i,X^k]E_{kj}\, , 
 \end{equation}
and CS action given by
 \begin{equation}
 \label{di}
 S^{\rm CS}=T_0 \int d\tau {\rm STr} \{ -P[k^{-2} k^{(1)}]+iP[(\incl_X \incl_X)C^{(3)}] +
 \frac12 P[(\incl_X\incl_X)^2\incl_k C^{(6)}] -\frac{i}{6} P[(\incl_X\incl_X)^3
 \incl_k N^{(8)}]\}
 \end{equation}
 where $\incl_k N^{(8)}$ denotes the Kaluza-Klein monopole potential \cite{BEL}.
 In this action
 $k^\mu$ is an Abelian Killing vector that points on the direction of propagation of the
 waves. This direction is isometric, because the background fields are
 either contracted with the Killing vector, so that any component along the isometric
 direction of the contracted field vanishes, or pulled back in the worldvolume with
 covariant derivatives relative to the isometry (see \cite{JL2} for their explicit 
 definition)\footnote{The reduced metric ${\cal G}_{\mu\nu}$ appearing in (\ref{mred})
 is in fact defined such that its pull-back with ordinary derivatives equals the pull-back
 of $g_{\mu\nu}$ with these covariant derivatives.}.
  To understand why this is so we need to recall the construction of this action. 

Expression (\ref{Mwaves}) was obtained by uplifting to eleven dimensions the action for Type 
IIA gravitational waves derived  in \cite{JL1} using Matrix String theory in a weakly curved background, 
and then going beyond the weakly curved background approximation
 by demanding agreement with Myers action for D0-branes 
when the waves propagate along the eleventh direction. 

In the action for Type IIA waves the 
circle in which one Matrix theory is compactified in order to construct Matrix String theory
cannot be 
decompactified in the non-Abelian case \cite{JL1}. In fact, the action exhibits a 
$U(1)$ isometry associated to translations along this direction, which by construction is
also the direction on which the waves propagate.
A simple way to see this is to recall that the 
last operation in the 9-11 flip involved in the construction of Matrix String theory
is a T-duality from fundamental strings wound around the 
9th direction. Accordingly, in the action we find a minimal coupling to 
$g_{\mu 9}/g_{99}$, which is the momentum operator $k^{-2}k_\mu$ 
in adapted coordinates. Therefore, by 
construction, the action (\ref{Mwaves}) is designed to describe BPS waves with 
momentum charge along the compact isometric direction.
 It is important to mention that in the Abelian limit, when all dielectric couplings and $U(N)$ covariant
 derivatives\footnote{Which are of course implicit in the pull-backs of the non-Abelian
 action (\ref{Mwaves}).}
  disappear, (\ref{Mwaves}) can be Legendre transformed into an action in
 which the dependence on the isometric direction has been restored. This action is
 precisely the usual action for a massless particle written in terms of an auxiliary $\gamma$  
 metric (see \cite{JL2} and \cite{JL1} for the details), where no information remains about the 
 momentum charge carried by the particle.  
 
 Let us now look at the couplings to the 3-form potential of eleven dimensional supergravity. We clearly find a dipole coupling in the 
 CS part of the action and a
 magnetic moment coupling
 \begin{equation}
 \label{mm}
 [X^i,X^k](\incl_k C^{(3)})_{kj}
 \end{equation}
 in the BI part\footnote{Let us stress that through its contraction with the Killing vector the 3-form potential acquires the necessary rank to couple in the BI action.}.
 These couplings play a crucial role
 in the microscopical description of the $AdS_4\times S^7$ dual giant graviton and the $AdS_7\times S^4$ giant graviton, respectively. 
 
 Let us parametrise 
 the $AdS_m\times S^n$ background as
 \begin{equation}
 \label{ads}
 ds^2=-(1+\frac{r^2}{{\tilde  L}^2})dt^2+\frac{dr^2}{1+\frac{r^2}{{\tilde L}^2}}+r^2d\Omega_{m-2}^2
 +L^2(d\theta^2+\cos^2{\theta}d\phi^2+\sin^2{\theta}d\Omega_{n-2}^2)
 \end{equation}
 \begin{equation}
 C^{(m-1)}_{t\alpha_1\dots\alpha_{m-2}}=-\frac{r^{m-1}}{{\tilde L}}\sqrt{g_\alpha}\, ,\qquad
 C^{(n-1)}_{\phi\beta_1\dots\beta_{n-2}}=a_n L^{n-1}\sin^{n-1}{\theta}\sqrt{g_\beta}
 \end{equation}
 where $a_4=a_5=1$,  $a_7=-1$, $\alpha_i$ ($\beta_i$) parametrise the
 $S^{m-2}$ ($S^{n-2}$) contained in $AdS_m$ ($S^n$)  as
 \begin{equation}
 d\Omega^2_{m-2}=d\alpha_1^2+\sin^2{\alpha_1}(d\alpha_2^2+\sin^2{\alpha_2}
 (\dots +\sin^2{\alpha_{m-3}} d\alpha^2_{m-2}))
 \end{equation}
 (similarly for $\beta_i$) and $\sqrt{g_\alpha}$ ($\sqrt{g_\beta}$)
 denotes the volume element on the unit $S^{m-2}$ ($S^{n-2}$).
 
 Consider the $AdS_7\times S^4$ background and take $r=0$, $\theta={\rm constant}$ 
 and $\phi$ time-dependent, i.e. the ansatz  for the giant graviton configuration. In this case there is a non-vanishing 3-form potential
 \begin{equation}
 C^{(3)}_{\phi\beta_1\beta_2}=L^3\sin^3{\theta}\sqrt{g_\beta}
 \end{equation}
 which (when rewritten in terms of Cartesian coordinates) clearly couples as in (\ref{mm}), given that the gravitons carry $P_\phi$
 angular momentum,  which identifies $\phi$ with the isometric direction in the action (so that $k^\mu=\delta^\mu_\phi$).
 On the other hand, taking the $AdS_4\times S^7$ background and the dual giant graviton
 ansatz, $\theta=0$, $r={\rm constant}$ and $\phi$ time-dependent, the non-vanishing
 3-form potential is
 \begin{equation}
 C^{(3)}_{t\alpha_1\alpha_2}=-\frac{r^3}{{\tilde  L}}\sqrt{g_\alpha}
 \end{equation}
 which couples through (\ref{di}).
 The detailed computations of the potentials associated to these configurations were performed in \cite{JL2}, and perfect agreement was found for large number of gravitons with the 
 macroscopical calculations in \cite{GST,GMT}.
 
 Consider now the backgrounds in which the gravitons expand into M5-branes.
 Taking the giant graviton ansatz in the $AdS_4\times S^7$ background one finds a 
 non-vanishing 6-form potential
 \begin{equation}
 \label{6mag}
C^{(6)}_{\phi\beta_1\dots\beta_5}=-L^6\sin^6{\theta}\sqrt{g_\beta}\, .
\end{equation}
Clearly this potential does not couple in the action for the system of waves if we identify $\phi$ with the isometric direction, given that in the pull-back involved in 
\begin{equation}
\label{C6}
\int d\tau P[(\incl_X\incl_X)^2 \incl_k C^{(6)}]
\end{equation}
only $\phi$ is time-dependent, and this component is already taken through the interior product with $k^\mu$.
Similarly, the dual giant graviton
ansatz in $AdS_7\times S^4$ yields 
\begin{equation}
\label{6di}
C^{(6)}_{t\alpha_1\dots\alpha_5}=-\frac{r^6}{{\tilde  L}}\sqrt{g_\alpha}
\end{equation}
which again does not couple in the action, this time because the quadrupolar coupling (\ref{C6})
has $\phi$ component and therefore is not of the form (\ref{6di}).

A puzzle then arises regarding the microscopical description of these giant graviton configurations
in terms of dielectric gravitational waves. 
By analogy with other expanded configurations one would expect the gravitons to expand into fuzzy 5-spheres due to quadrupolar electric or magnetic moment couplings to the 6-form potential. Since a 5-sphere has 5 relative dimensions with respect to a point-like object, the 6-form potential has to be contracted as well with the Killing direction in order to be able to couple to a one dimensional
worldvolume. 
However, we have seen that if this Killing direction is the direction of propagation the only coupling of this form present in the action for M-theory waves (\ref{Mwaves}) vanishes for the $AdS$ backgrounds that we want to study.

\vspace{1cm}

In order to find a possible solution to this puzzle let us recall the microscopical description
of the giant graviton configurations in the $AdS_5\times S^5$ background \cite{JLR}.
These configurations correspond microscopically
to Type IIB gravitational waves expanding into $S^3$ D3-branes due to their 
dielectric or magnetic moment interaction with the 4-form 
RR potential of the background. This potential is
\begin{equation}
\label{c4}
C^{(4)}_{\phi\beta_1\beta_2\beta_3}=L^4\sin^4{\theta}\sqrt{g_\beta}\, ,
\end{equation}
for the giant graviton, and
\begin{equation}
\label{c4d}
C^{(4)}_{t\alpha_1\alpha_2\alpha_3}=-\frac{r^4}{L}\sqrt{g_\alpha}
\end{equation}
for the dual giant graviton.
Consider now that the action for Type IIB waves contained the coupling that one would
naturally expect to involve the 4-form potential:
\begin{equation}
\int d\tau P[(\incl_X\incl_X)i_k C^{(4)}]\, .
\end{equation}
By the same arguments above one easily checks that this coupling vanishes
both for the giant and dual giant graviton potentials.

The action describing Type IIB waves constructed in \cite{JLR}
contains however a second isometric direction, with Killing vector 
$l^\mu$.
This direction is the direction in which one performs the T-duality
transformation that has to be made in order to obtain the action for Type IIB waves 
from the (reduction of the) action for M-theory waves. 
In the Abelian limit one can interpret the resulting action as a dimensional
reduction over the T-duality direction, as one usually does, but this cannot be done in 
the non-Abelian
case, in part due to the presence of non-trivial dielectric couplings. One finds, in particular,
the following coupling in the CS part of the action
\begin{equation}
\label{C4di}
\int d\tau P[(\incl_X\incl_X)i_l C^{(4)}]\, ,
\end{equation}
together with a magnetic moment coupling to the 4-form potential in the 
BI part of the action, 
\begin{equation}
\label{C4mag}
[X^i,X^k] (i_k i_l C^{(4)})_{kj}\, .
\end{equation}

This isometric action for Type IIB waves
is therefore valid to describe waves propagating in 
backgrounds which contain a $U(1)$ isometric
direction. This is the case for the
$AdS_5\times S^5$ background, where the $U(1)$ isometry is that corresponding
to the translations along the
$S^1$-fibre in the description of the 3-sphere (contained in
$S^5$ (for the giant graviton) or $AdS_5$ (for the dual giant graviton)) as an
$S^1$-fibre over an $S^2$ base manifold. 
In fact rewriting the potentials (\ref{c4}) and (\ref{c4d})
in adapted coordinates to this isometry it is easy to see that the coupling (\ref{C4di})
is non-vanishing for the 4-form potential associated to the dual giant graviton 
and the coupling (\ref{C4mag}) for the one associated to the giant graviton. Indeed the 
detailed computation of the corresponding non-Abelian potentials shows perfect agreement with
the macroscopical calculations in \cite{GST,HHI} for large number of gravitons.
Let us stress however that the right dielectric couplings that cause the expansion
of the gravitons can only be constructed in spacetimes with a $U(1)$ isometry. 

\vspace{1cm}

The discussion above suggests that something similar can be happening
for the gravitons expanding into 5-spheres that we are considering in this
article.
The $S^5$ can similarly be described as a $U(1)$ bundle, in this case
over the two dimensional complex projective space, $CP^2$. Therefore, there is 
a $U(1)$ isometry in the background that could allow the
construction of further dielectric couplings. 

However, the action
that we know for M-theory waves contains only one isometric direction, that we naturally
identified
with the direction of propagation of the waves, since, as we discussed,
they are minimally coupled to the
momentum operator in this direction. 
Consider instead that we identified this direction
with the $U(1)$ fibre of the $U(1)$-decomposition of the 5-sphere.
In this case one can see that the coupling (\ref{C6}) is non-vanishing both for the giant and dual giant graviton potentials. We will see this in detail in the next sections, when we write $S^5$ as an $S^1$ fibre over $CP^2$ in adapted coordinates. Denoting $\chi$ the coordinate adapted to the isometry, (\ref{C6}) becomes
\begin{equation}
\label{C6bis}
\int d\tau X^l X^k X^j X^i \Bigl[C^{(6)}_{\chi ijkl 0}+C^{(6)}_{\chi ijkl \phi}\dot{\phi}\Bigr]\, ,
\end{equation}
and one easily finds that the first term is non-vanishing for the dual giant graviton whereas the second one is non-vanishing for the giant graviton.

We then propose to use the action (\ref{Mwaves}) to describe gravitons propagating in a 
spacetime with a compact isometric direction, $\chi$, with, by construction, 
non-vanishing momentum-charge along that direction, $P_\chi$, and with a non-zero velocity 
along a different transverse direction, $\dot{\phi}$, as implied by the giant 
and dual giant gravitons ans\"atze. 
Clearly, in order to describe giant graviton configurations, which only carry momentum $P_\phi$, 
we will have to set  the momentum charge $P_\chi$ to zero at the end of the calculation. We will see that this calculation matches exactly the macroscopical calculation in \cite{GST,GMT} for large number of gravitons, which will provide the strongest check for the validity of our proposal.

Of course, a more direct microscopical description of giant gravitons with momentum 
$P_\phi$ living in a spacetime with an isometric direction $\chi$ would be in terms 
of an effective action containing both $\chi$ and $\phi$ as isometric directions, 
and with momentum charge only with respect to the second one.
As we have seen this type of action exists in the Type IIB theory. However the construction 
of a similar type of action for M-theory waves cannot be made based on duality arguments.
One possibility would be to start from the two isometric action for Type IIB waves, T-dualize and
uplift to eleven dimensions. The detailed calculation shows that in the non-Abelian case
both the T-duality direction and the eleventh dimension become isometric directions in
the M-theory action. Therefore, the resulting action is adequate for the study of M-theory
gravitational waves propagating in a spacetime with three $U(1)$ directions (the other
isometric direction is the direction of propagation of the waves). These isometries are
however not present in the M-theory backgrounds that we want to study.

Let us finally remark that an action for M-theory gravitational waves with two isometric
directions would have to be highly non-perturbative. This would
be in the same spirit of \cite{BMN}, where it is argued that the 5-brane cannot appear 
as a classical solution to the pp-wave Matrix model because the scaling of its radius
with the coupling constant is more non-perturbative than
the one corresponding to a classical solution.
Coming back to our action, if the direction of propagation is also isometric, 
we cannot have a magnetic coupling to 
the 6-form potential like the second one in (\ref{C6bis}). Therefore the
only way to find such a magnetic coupling is in the BI action, through something like 
\begin{equation}
Q^i_j=\delta^i_j +\dots +[X^i,X^k][X^l,X^m](\incl_k\incl_l C^{(6)})_{klmj}\, .
\end{equation}
However, quadratic couplings of this sort are by no means predicted 
by T-duality (plus the uplift to M-theory)\footnote{This is not the case for the Type IIB waves, where the corresponding coupling is a dipole coupling
$$Q^i_j=\delta^i_j+\dots-i[X^i,X^k](\incl_k\incl_l C^{(4)})_{kj}$$
neatly predicted by T-duality from the action for Type IIA waves (see \cite{JLR} for the details).}.
Clearly this is due to the fact that an action containing this kind of couplings must be highly 
non-perturbative. Therefore it could only be derived \`a la Myers from a non-perturbative action
for coincident D-branes. Although such a non-perturbative action is known for a single brane 
(it is well-known that worldvolume duality of the BI vector yields the action which is valid 
in the strong coupling regime) it is not known for coincident branes.
Therefore, we seem to be stuck with some fundamental problem in D-brane actions.

In the next two sections we show how the action (\ref{Mwaves}) can be used to correctly 
describe microscopically the giant graviton in $AdS_4\times S^7$ and the dual giant 
graviton in $AdS_7\times S^4$. We will compare in both cases to the corresponding 
macroscopical descriptions and see that there is perfect agreement for large number of gravitons.

\sect{The giant graviton in $AdS_4 \times S^7$}

The giant graviton solution of \cite{GST} in this spacetime is in terms of an
M5-brane with the topology of a 5-sphere contained in $S^7$,
carrying angular momentum along the $\phi$-direction and magnetic moment with respect
to the 6-form potential of the background:

\begin{equation}
C^{(6)}_{\phi\beta_1\dots\beta_5}=-L^6\sin^6{\theta}\sqrt{g_\beta}
\end{equation}
where $\beta_i$ are the angles parametrising the unit 5-sphere in $S^7$:

\begin{equation}
d\Omega_5^2=d\beta_1^2+\sin^2{\beta_1}(d\beta_2^2+\sin^2{\beta_2}(d\beta_3^2+
\sin^2{\beta_3}(d\beta_4^2+\sin^2{\beta_4}d\beta_5^2)))
\end{equation}

Taking the ansatz $r=0$, $\theta=$ constant, $\phi=\phi(\tau)$ in (\ref{ads}) we expect to find the gravitons
expanding into a non-conmutative $S^5$ with radius $L\sin{\theta}$. As we have mentioned in the previous section, describing the $S^5$
as a $U(1)$ bundle over the two dimensional complex projective space, $CP^2$, it is natural to identify
the isometry of the action for the coincident gravitons with the $U(1)$ coordinate.

\subsection{$S^5$ as a $U(1)$-bundle over $CP^2$}

It is well-known that the 5-sphere can be described as a $U(1)$ bundle over the two
dimensional complex projective space. Here we will review some details of this
construction and introduce a convenient set of adapted coordinates to the $U(1)$
isometry. We will mainly follow references \cite{GP,pope}. The reader is referred to
those references for more details.

The unit $S^5$ can be represented as a submanifold of
$\C^3$ with coordinates $(z_0,z_1,z_2)$ satisfying 
$\bar{z}_0 z_0+{\bar z}_1 z_1 +\bar{z}_2 z_2=1$. 
This is invariant under $z_i\rightarrow z_i e^{i\alpha}$, and $CP^2$ is the space of
orbits under the action of this circle group. The projection of points in $S^5$
onto these orbits is the $U(1)$-fibration of $S^5$. Setting:
\begin{equation}
\xi_1=z_1/z_0\, , \qquad \xi_2=z_2/z_0\, , \qquad z_0=|z_0|e^{i\chi}\, ,
\end{equation}
and defining:
\begin{equation}
\label{Acon}
A=\frac{i}{2}(1+|\xi_1|^2+|\xi_2|^2)^{-1}[\bar{\xi}_1 d\xi_1+\bar{\xi}_2 d\xi_2
-{\rm c.c.}]\, ,
\end{equation}
the metric on the $S^5$ may be written as
\begin{eqnarray}
\label{5sph}
d\Omega_5^2&=&(d\chi-A)^2+(1+\sum_k |\xi_k|^2)^{-1}\sum_i |d\xi_i|^2- \nonumber\\
&&-(1+\sum_k |\xi_k|^2)^{-2}\sum_{i,j}\xi_i\bar{\xi}_j d\bar{\xi}_id\xi_j \nonumber\\
&=&(d\chi-A)^2+ds^2_{CP^2}\, ,
\end{eqnarray}
since $CP^2$ is the projection orthogonal to the vector $\partial /\partial\chi$.
$ds^2_{CP^2}$ is the Fubini-Study metric for $CP^2$:
\begin{eqnarray}
ds^2_{CP^2}&=&(1+\sum_k |\xi_k|^2)^{-1}\sum_i |d\xi_i|^2
-(1+\sum_k |\xi_k|^2)^{-2}\sum_{i,j}\xi_i\bar{\xi}_j d\bar{\xi}_id\xi_j= \nonumber\\
&=&\frac{\partial^2 K}{\partial\xi^i\partial\bar{\xi}^{\bar{j}}}d\xi^i
d\bar{\xi}^{\bar{j}}\, ;
\end{eqnarray}
with $K=\log{(1+|\xi^1|^2+|\xi^2|^2)}$. Therefore $CP^2$ has a K\"ahler structure with
K\"ahler form $J=i\partial\bar{\partial}K$. The field strength $F=dA=2J$ is a 
solution of Maxwell's equations, the so-called ``electromagnetic instanton'' of 
\cite{Traut}. It is self-dual and satisfies that its integral $\int F\wedge F$ 
associated with the 
second Chern class is equal to $4\pi^2$. This solution to Maxwell's equations will in
fact play a role in the macroscopical description of giant gravitons
of sections 3.4 and 4.1.

One can obtain a real four dimensional metric on $CP^2$ by defining coordinates
$(\varphi_1,\varphi_2,\psi,\varphi_3)$, $0\leq\varphi_1\leq \pi/2$,
$0\leq\varphi_2\leq\pi$, $0\leq\psi\leq 4\pi$, $0\leq \varphi_3\leq 2\pi$, as
\cite{pope}:

\begin{eqnarray}
\xi_1&=&\tan{\varphi_1}\cos{\frac{\varphi_2}{2}}e^{i(\psi+\varphi_3)/2} \nonumber\\
\xi_2&=&\tan{\varphi_1}\sin{\frac{\varphi_2}{2}}e^{i(\psi-\varphi_3)/2}
\end{eqnarray}
to give
\begin{equation}
\label{phis}
ds^2_{CP^2}=d\varphi_1^2+\frac14 \sin^2{\varphi_1}\Bigl[\cos^2{\varphi_1}
(d\psi+\cos{\varphi_2}d\varphi_3)^2+d\varphi_2^2+
\sin^2{\varphi_2}d\varphi_3^2\Bigr]\, .
\end{equation}
In these coordinates the connection $A$ defined in (\ref{Acon}) is given by
\begin{equation}
\label{Acon2}
A=-\frac12 \sin^2{\varphi_1}(d\psi+\cos{\varphi_2} d\varphi_3)
\end{equation}
and the 6-form potential of the background reads:
\begin{equation}
C^{(6)}_{\phi\chi\varphi_1\varphi_2\psi\varphi_3}=-\frac18 L^6\sin^6{\theta}
\sin^3{\varphi_1}\sin{\varphi_2}\cos{\varphi_1}\, .
\end{equation}

Clearly, the background is isometric in the $\chi$-direction, and this is the direction that we are going to identify with the isometric direction in the action (\ref{Mwaves}), i.e. $k^\mu=\delta^{\mu}_\chi$.

Let us now make the non-commutative ansatz for the 5-sphere.
Inspired by the results of \cite{JLR} for $AdS_5\times S^5$, where it was found that
the non-commutative manifold on which the giant (and dual giant) graviton expands is
defined as an $S^1$ bundle over a non-commutative $S^2$, 
we make the ansatz that the non-commutative manifold onto
which the giant graviton in $AdS_4\times S^7$ 
expands is defined as an $S^1$ bundle over a non-commutative
$CP^2$. 

\subsection{The fuzzy $CP^2$}

In this subsection we review some basic properties about the fuzzy $CP^2$. The
fuzzy $CP^2$ has been extensively studied in the literature (see for instance
\cite{TV},\cite{NR}-\cite{KNR}). In this section we will mainly follow the notation in \cite{ABIY}.

$CP^2$ is the coset manifold $SU(3)/U(2)$, and can be defined as the submanifold of
$\R^8$ determined by the constraints:
\begin{eqnarray}
\label{condi}
&&\sum_{i=1}^8 x^i x^i=1 \nonumber\\
&&d^{ijk}x^j x^k =\frac{1}{\sqrt{3}}x^i
\end{eqnarray}
where $d^{ijk}$ are the components of the totally symmetric 
$SU(3)$-invariant tensor defined 
by
\begin{equation}
\lambda^i \lambda^j=\frac23 \delta^{ij}+(d^{ijk}+if^{ijk})\lambda^k\, ,
\end{equation}
where $\lambda^i$, $i=1,\dots,8$ are the Gell-Mann matrices.
In this set of constraints only four are independent (the first one, for instance,
is a consequence of the rest), therefore they define a four dimensional manifold.

A matrix level definition of the fuzzy $CP^2$
can be obtained by impossing the conditions (\ref{condi}) at the level of matrices. 
Defining a set of coordinates $X^i$, $i=1,\dots,8$ as
\begin{equation}
\label{defX}
X^i=\frac{1}{\sqrt{C_N}} T^i
\end{equation}
with $T^i$ the generators of $SU(3)$ in an $N$ dimensional irreducible representation and $C_N$ the
quadratic Casimir of $SU(3)$ in this representation,  the first 
constraint in (\ref{condi}) is trivially satisfied through the quadratic Casimir of the group
\begin{equation}
\sum_{i=1}^8 X^i X^i=\frac{1}{C_N}\sum_{i=1}^8 T^i T^i =\frac{1}{C_N}C_N=
\unity\, ,
\end{equation}
whereas the rest of the constraints are satisfied for any $n$ 
(see Appendix A in \cite{ABIY})
if the $X^i$'s are taken 
in the $(n,0)$ or
$(0,n)$ representations of $SU(3)$, parametrising the irreducible representations of 
$SU(3)$ by two integers
$(n,m)$ corresponding to the number of fundamental and anti-fundamental indices.
They become at the level of matrices
\begin{equation}
\label{const2}
d^{ijk}X^jX^k=\frac{\frac{n}{3}+\frac12}{\sqrt{\frac13 n^2+n}}X^i\, .
\end{equation}

Therefore, to describe the fuzzy $CP^2$ the non-commuting coordinates $X^i$ have to be
taken in the $(n,0)$ or $(0,n)$ irreducible representations of $SU(3)$.
This is in agreement with the fact that $G/H$ cosets can be made fuzzy if $H$ is the
isotropy group of the lowest weight state of a given irreducible representation of $G$
\cite{Madore,TV}. Therefore,
different irreducible representations, having associated different isotropy
subgroups, can give rise to different cosets $G/H$.  $CP^2$ has $G=SU(3)$ and $H=U(2)$, and
this is precisely  the isotropy subgroup of the $SU(3)$ irreducible representations $(n,0)$ and $(0,n)$. 
Any other choice of $(n,m)$ has isotropy group $U(1)\times U(1)$, and therefore yields
to a different coset, $SU(3)/(U(1)\times U(1))$.

It will be useful later to know 
that the irreducible representations $(n,0)$, $(0,n)$ have dimension N
given by
\begin{equation}
N=\frac12 (n+1)(n+2)
\end{equation}
and quadratic Casimir
\begin{equation} 
C_N=\frac13 n^2+n\, .
\end{equation}

$CP^2$ can be embedded in $\R^8$ using coherent state techniques \cite{perelo} 
(see also Appendix B in \cite{TV}). In our coordinates (\ref{phis}) we have:
\begin{eqnarray}
X^1&=&\frac12 \sin{2\varphi_1}\cos{\frac{\varphi_2}{2}}
\cos{\frac{\psi+\varphi_3}{2}} \nonumber\\
X^2&=&-\frac12 \sin{2\varphi_1}\cos{\frac{\varphi_2}{2}}
\sin{\frac{\psi+\varphi_3}{2}} \nonumber\\
X^3&=&\frac12[\sin^2{\varphi_1}(1+\cos^2{\frac{\varphi_2}{2}})-1]\nonumber\\
X^4&=&\frac12 \sin{2\varphi_1}\sin{\frac{\varphi_2}{2}}
\cos{\frac{\psi-\varphi_3}{2}} \nonumber\\
X^5&=&-\frac12 \sin{2\varphi_1}\sin{\frac{\varphi_2}{2}}
\sin{\frac{\psi-\varphi_3}{2}} \nonumber\\
X^6&=&\frac12 \sin^2{\varphi_1}\sin{\varphi_2}\cos{\varphi_3} \nonumber\\
X^7&=&-\frac12 \sin^2{\varphi_1}\sin{\varphi_2}\sin{\varphi_3} \nonumber\\
X^8&=&\frac{1}{2\sqrt{3}}(3\sin^2{\varphi_1}\sin^2{\frac{\varphi_2}{2}}-1)\, ,
\end{eqnarray}
for which
\begin{equation}
\sum_{i=1}^8 (dX^i)^2=d\varphi_1^2+\frac14 \sin^2{\varphi_1}\Bigl[\cos^2{\varphi_1}
(d\psi+\cos{\varphi_2}d\varphi_3)^2+d\varphi_2^2+
\sin^2{\varphi_2}d\varphi_3^2\Bigr]=ds^2_{CP^2}\, .
\end{equation}
We also have that
\begin{equation}
\sum_{i=1}^8 (X^i)^2=\frac13\, ,
\end{equation}
so that in order to fulfill this constraint 
we will have to slightly modify
our definition (\ref{defX}). We then get, in the $(n,0)$ or $(0,n)$ representations
\begin{equation}
X^i=\frac{1}{\sqrt{3}\sqrt{C_N}}T^i=\frac{1}{\sqrt{n^2+3n}}T^i\, , \qquad
i=1,\dots,8\, .
\end{equation}

With this normalisation the commutation relations of the $X^i$ become
\begin{equation}
[X^i,X^j]=\frac{i}{\sqrt{n^2+3n}}f^{ijk}X^k\, ,
\end{equation}
with $f^{ijk}$  the structure constants of $SU(3)$ in the notation
\begin{equation}
[\lambda^i,\lambda^j]=2i f^{ijk} \lambda^k\, .
\end{equation}

\subsection{The microscopical description}

Let us now take the giant graviton ansatz, 
$r=0$, $\theta={\rm constant}$, $\phi=\phi(\tau)$ in the
$AdS_4\times S^7$ background. We find, in Cartesian coordinates
\begin{equation}
ds^2=-dt^2+L^2\cos^2{\theta}d\phi^2+L^2\sin^2{\theta}\Bigl[(d\chi-A)^2
+(dX^1)^2+\dots+(dX^8)^2\Bigr]
\end{equation}
and
\begin{equation}
C^{(6)}_{\phi\chi ijkl}=2L^6\sin^6{\theta} f^{[ijm}f^{kl]n}X^mX^n\, . 
\end{equation}

Taking now $k^\mu=\delta^\mu_\chi$ in the action (\ref{Mwaves}) we have
that
\begin{eqnarray}
&&k=L\sin{\theta}\, , \qquad E_{00}=-1+L^2\cos^2{\theta}{\dot{\phi}}^2\, , \nonumber\\
&&Q^i_j=\delta^i_j-\frac{L^3\sin^3{\theta}}{\sqrt{n^2+3n}}f^{ijk} X^k\, ,
\qquad i,j=1,\dots,8\, ,
\end{eqnarray}
and substituting in the action we find
\begin{equation}
S^{\rm BI}=-T_0\int d\tau {\rm STr}\Bigl\{\frac{1}{L\sin{\theta}}\sqrt{1-L^2\cos^2{\theta}
{\dot{\phi}}^2}
\sqrt{\unity+\frac32 \frac{L^6\sin^6{\theta}}{n^2+3n}X^2+\frac{9}{16}
\frac{L^{12}\sin^{12}{\theta}}{(n^2+3n)^2}X^2X^2+\dots}\Bigr\}\, .
\end{equation}
Here we have dropped those contributions to ${\rm det}Q$ that will vanish when taking
the symmetrised trace, and ignored higher powers of $n^2+3n$ which will vanish
in the large $N$ ($\Leftrightarrow n\rightarrow \infty$) limit.
These terms on the other hand cannot be nicely arranged into higher powers of the
quadratic Casimir without explicit use of the constraints (\ref{const2}).
Up to order $n^{-4}$ we have that
\begin{equation}
\sqrt{\unity+\frac32 \frac{L^6\sin^6{\theta}}{n^2+3n}X^2+\frac{9}{16}
\frac{L^{12}\sin^{12}{\theta}}{(n^2+3n)^2}X^2X^2+\dots}=\unity+\frac34 
\frac{L^6\sin^6{\theta}}{n^2+3n}X^2\, .
\end{equation}

We also have for the CS part of the action:
\begin{equation}
S^{\rm CS}=\frac{T_0}{2}\int d\tau {\rm STr}\Bigl\{P[(\incl_X\incl_X)^2
i_k C^{(6)}]\Bigr\}=\int d\tau \frac{NT_0}{4}\frac{L^6\sin^6{\theta}}{n^2+3n}\dot{\phi}
\end{equation}

It is then easy to compute the symmetrised trace to finally
arrive, in Hamiltonian
formalism, to
\begin{equation}
\label{Hgiant}
H=\frac{P_\phi}{L}\sqrt{1+\tan^2{\theta}\Bigl(1-\frac{NT_0}{4(n^2+3n)P_\phi}
L^6\sin^4{\theta}\Bigr)^2+\frac{N^2T_0^2}{P_\phi^2\sin^2{\theta}}\Bigl(1+\frac12
\frac{L^6\sin^6{\theta}}{n^2+3n}\Bigr)}
\end{equation}
where $P_\phi$ is a conserved quantity given that $\phi$ is cyclic in the Lagrangian.
We should stress that this expression is an approximated expression in which higher
powers of $n^2+3n$ vanishing in the large $n$ limit have been omitted.

In order to describe giant graviton configurations we must set to zero the momentum
along the $\chi$-direction, $P_\chi$. Recall however that, by construction, the $N$
gravitons carry momentum $P_\chi=NT_0$.
The difference between $P_\chi$ being zero or not is merely
a coordinate transformation, a boost in $\chi$. However, how to perform coordinate 
transformations in non-Abelian actions is an open problem \cite{DS, Hassan, DSW, Dom}.
In order to clarify how the limit $P_\chi\rightarrow 0$ must be taken we study
in the next subsection the macroscopical description of this microscopical 
configuration, in terms of a spherical M5-brane carrying both $P_\phi$ and
$P_\chi$ charges. The agreement between the two descriptions in the large $N$ limit
will make
clear that in the limit $P_\chi\rightarrow 0$ 
the last term in the Hamiltonian should vanish. On the other hand, the term
\begin{equation}
\label{nv}
\frac{NT_0}{4(n^2+3n)P_\phi}\, ,
\end{equation}
whose presence is in fact crucial in
order to find the giant graviton configuration, remains finite, since
$N=(n+1)(n+2)/2$ for $(n,0)$ and $(0,n)$ irreps, so that
numerator and denominator in (\ref{nv}) scale with the same power of $n$. 
In fact, in the $N\rightarrow \infty$ ($\Leftrightarrow n\rightarrow\infty$) limit both terms are compensated, and the finite result
$T_0/8P_\phi$ is reached, in perfect agreement with the macroscopical calculation.
This compensation does not however take place in the last term in the Hamiltonian.

Therefore, in order to describe giant graviton configurations we minimize
\begin{equation}
\label{Hmic}
H_{\rm mic}=\frac{P_\phi}{L}\sqrt{1+\tan^2{\theta}\Bigl(1-\frac{NT_0}{4(n^2+3n)P_\phi}
L^6\sin^4{\theta}\Bigr)^2}
\end{equation}
with respect to $\theta$. We then find two solutions with energy $P_\phi/L$: $\sin{\theta}=0$, 
which corresponds to the point-like
graviton, and 
\begin{equation}
\sin{\theta}=\Bigl(\frac{4(n^2+3n)P_\phi}{NT_0L^6}\Bigr)^{1/4}
\end{equation}
which corresponds to the giant graviton, with radius
\begin{equation}
R=\Bigl(\frac{4(n^2+3n)P_\phi}{NT_0L^2}\Bigr)^{1/4}\, .
\end{equation}
Clearly, for the giant graviton  
\begin{equation}
P_\phi\leq\frac{NT_0L^6}{4(n^2+3n)}
\end{equation}
which provides the microscopical bound to the angular momentum predicted by the
stringy exclusion principle. When the number of gravitons is large we find that
\begin{equation}
P_\phi\leq\frac{T_0L^6}{8}={\tilde N}
\end{equation}
with ${\tilde N}$ the units of 6-form flux on the 5-sphere in the macroscopical
description of \cite{GST,GMT}, given by
${\tilde N}=A_5 T_5 L^6$ with $A_5$ the area of the unit 5-sphere. We therefore
find perfect agreement with the bound found in \cite{GST}. The same holds true for the
radius of the configuration, which for large number of gravitons is given by
\begin{equation}
R=\Bigl(\frac{8P_\phi}{T_0L^2}\Bigr)^{1/4}=L 
\Bigl(\frac{P_\phi}{\tilde N}\Bigr)^{1/4}
\end{equation}
exactly as in \cite{GST,GMT}.

In this section we have achieved the right microscopical description of the giant graviton in $AdS_4\times S^7$ in
terms of gravitons expanding into a ``fuzzy 5-sphere'', defined as a $U(1)$ bundle over the
fuzzy $CP^2$. We have seen that the coordinate along the fibre must be isometric in the action, 
and this has forced our choice of $k^\mu$ pointing on that direction, which has in turn
introduced, by construction, a non-vanishing momentum $P_\chi$ on the configuration. The 
macroscopical description of such a configuration should then be in terms of a spherical 
M5-brane carrying both
$P_\phi$ and $P_\chi$ charges. We will perform the detailed macroscopical description in these
terms in the next subsection. The agreement between the two descriptions will provide 
further justification to the limit taken to arrive at expression (\ref{Hmic}) as the 
right microscopical Hamiltonian  describing gravitons with angular momentum $P_\phi$.

\subsection{The macroscopical calculation}

The simplest way to describe an M5-brane living in a spacetime with a
$U(1)$ isometry and carrying momentum along that direction is by uplifting
a D4-brane to M-theory keeping the eleventh direction compact. The resulting
brane is therefore a longitudinal 5-brane. Doing this we
obtain the action adequate to describe an M5-brane whose worldvolume
contains an isometric direction. Clearly this applies to the M5-brane with the
topology of an $S^5$. In this case the isometric direction is the coordinate along the
fibre in the decomposition of the $S^5$ as a $U(1)$ fibre over $CP^2$.
The worldvolume of such a brane is therefore effectively four-dimensional,
and locally that of a $CP^2$.

For the non-vanishing eleven dimensional fields involved in our
$AdS$ backgrounds we find the following action for a wrapped M5-brane:
\begin{equation}
\label{M5w}
S=-T_4 \int d^5\xi \Bigl\{ k \sqrt{{\rm det}({\cal G}+k^{-1}F)}-P[i_k C^{(6)}]
-\frac12 P[k^{-2} k^{(1)}]\wedge F\wedge F\Bigr\}\, .
\end{equation}
Here $F$ is the field strength associated to M2-branes wrapped on the isometric direction ending on the
M5-brane, since the uplifting of the BI field strength of the D4-brane, $F+B^{(2)}$, 
to M-theory gives
$F+i_k C^{(3)}$ \footnote{In the action (\ref{M5w}) we have set $C^{(3)}$ to zero,
which is valid for our particular backgrounds.}. ${\cal G}$ is the reduced metric defined in 
(\ref{mred}) and
we have denoted $T_4$ the tension of the brane to explicitly take into account that
its spatial worldvolume is 4-dimensional. 

As we discussed in Section 2 $k^{-2}k^{(1)}$ is identified with the momentum operator
along the isometric direction. Therefore we can switch on momentum charge
on the M5-brane by choosing a non-vanishing field strength such that
\begin{equation}
\label{FF}
\int_{CP^2}F\wedge F=8\pi^2 N\, ,
\end{equation}
since then
\begin{equation}
\frac{T_4}{2}\int d\tau d^4\sigma P[k^{-2}k^{(1)}]\wedge F\wedge F=
NT_0\int d\tau P[k^{-2}k^{(1)}]\, .
\end{equation}
With $F$ satisfying (\ref{FF}) we are therefore dissolving $N$ gravitons, propagating 
along the isometric direction, in the worldvolume of the 5-brane.
The field satisfying this condition is in fact the electromagnetic instanton that we
discussed in section 3.1, which now must have instanton number 
equal to twice the number of gravitons. This gives
\begin{equation}
F=\sqrt{\frac{N}{2}}\Bigl(-\sin{2\varphi_1}d\varphi_1\wedge d\psi-
\sin{2\varphi_1}\cos{\varphi_2}d\varphi_1\wedge d\varphi_3+
\sin^2{\varphi_1}\sin{\varphi_2}d\varphi_2\wedge d\varphi_3\Bigr)\, .
\end{equation}

Identifying the isometric direction in the action (\ref{M5w}) with $\chi$ in (\ref{5sph})
and integrating over the spatial worldvolume of the M5-brane we arrive at
\begin{equation}
S=-4\pi^2T_4\int d\tau \Bigl\{ L\sin{\theta}\sqrt{1-L^2\cos^2{\theta}{\dot{\phi}}^2}
\Bigl(\frac{L^4\sin^4{\theta}}{8}+\frac{N}{L^2\sin^2{\theta}}\Bigr)-\frac18 L^6\sin^6{\theta}
\dot{\phi}\Bigr\}
\end{equation}
and, in Hamiltonian formalism, to
\begin{equation}
\label{Hgiantm}
H=\frac{P_\phi}{L}\sqrt{1+\tan^2{\theta}\Bigl(1-\frac{\pi^2T_4L^6\sin^4{\theta}}
{2P_\phi}\Bigr)^2+\frac{16\pi^4T_4^2N^2}{P_\phi^2\sin^2{\theta}}
\Bigl(1+\frac{L^6\sin^6{\theta}}
{4N}\Bigr)}\, .
\end{equation}
Comparing with the Hamiltonian constructed in \cite{GST,GMT}, which describes a
spherical 5-brane with momentum $P_\phi$, we see that a non-vanishing momentum
along the $\chi$-direction translates into an additional piece depending on $N$
inside the squared root. Comparing
this expression with the microscopical Hamiltonian (\ref{Hgiant}) we find that they exactly agree in the large $N$ limit,
when $N\sim n^2/2$. In the macroscopical Hamiltonian (\ref{Hgiantm}) it is clearer however that the
limit of zero momentum along the fibre direction is reached when there are zero gravitons 
propagating in the $\chi$ direction
dissolved in the worldvolume of the 5-brane, which neatly sets to zero the last term in the
Hamiltonian. This further justifies the elimination of the corresponding term in
the microscopical Hamiltonian (\ref{Hgiant}) in section 3.3.

\section{The dual giant graviton in $AdS_7\times S^4$}

\subsection{The microscopical description}

In this section we briefly describe the microscopical description of the
dual giant graviton in $AdS_7\times S^4$. One expects that microscopically the
gravitons expand into a 5-sphere with radius $r$ due to the coupling to the
6-form potential $C^{(6)}_{t\alpha_1\dots\alpha_5}$. As before, we describe the
``fuzzy 5-sphere'' as a $U(1)$ bundle over the fuzzy $CP^2$, and we embed the $CP^2$ in 
$\R^8$. We then have for the dual giant graviton ansatz:
\begin{eqnarray}
&&ds^2=-(1+\frac{r^2}{{\tilde L}^2})dt^2+L^2d\phi^2+r^2\Bigl[(d\chi-A)^2+(dX^1)^2+\dots
(dX^8)^2\Bigr]\nonumber\\
&&C^{(6)}_{0\chi ijkl}=2\frac{r^6}{{\tilde L}}f^{[ijm}f^{kl]n}X^mX^n
\end{eqnarray}

Substituting in the action (\ref{Mwaves}) we have that
\begin{eqnarray}
&&k=r\, ,\qquad E_{00}=-(1+\frac{r^2}{{\tilde L}^2})+L^2{\dot{\phi}}^2\, ,\nonumber\\
&&Q^i_j=\delta^i_j-\frac{r^3}{\sqrt{n^2+3n}}f^{ijk}X^k\, , \qquad i,j=1,\dots,8
\end{eqnarray}
and
\begin{eqnarray}
S&=&-T_0\int d\tau {\rm STr}\Bigl\{\frac{1}{r}\sqrt{1+\frac{r^2}{{\tilde L}^2}-L^2{\dot{\phi}}^2}
\sqrt{\unity+\frac32 \frac{r^6}{n^2+3n}X^2+\frac{9}{16}\frac{r^{12}}{(n^2+3n)^2}X^2X^2
+\dots}\nonumber\\
&&-\frac{r^6}{{\tilde L}}\frac{N}{4(n^2+3n)}\Bigr\}\, .
\end{eqnarray}
Up to order $n^{-4}$ we arrive at the Hamiltonian
\begin{equation}
\label{Hgiantd}
H=\sqrt{\Bigl(1+\frac{r^2}{{\tilde L}^2}\Bigr)\Bigl(\frac{{P_\phi}^2}{L^2}
+\frac{N^2T_0^2r^{10}}{16(n^2+3n)^2}+\frac{N^2T_0^2}{r^2}
(1+\frac12 \frac{r^6}{n^2+3n})\Bigr)}-\frac{NT_0}{4(n^2+3n)}\frac{r^6}{{\tilde L}}
\end{equation}
Now we have to take the limit $P_\chi=0$ which amounts to setting to zero the last
term in the squared-root, as we discussed in detail when we studied the giant
graviton in $AdS_4\times S^7$. We are then left with
\begin{equation}
\label{Hmicd}
H_{\rm mic}=\sqrt{\Bigl(1+\frac{r^2}{{\tilde L}^2}\Bigr)\Bigl(\frac{{P_\phi}^2}{L^2}
+\frac{N^2T_0^2r^{10}}{16(n^2+3n)^2}\Bigr)}
-\frac{NT_0}{4(n^2+3n)}\frac{r^6}{{\tilde L}}\, ,
\end{equation}
which again is an approximated expression in which higher order powers of
$n^2+3n$ vanishing in the large $n$ limit have been omitted.
Minimizing with respect to $r$ we find two solutions with energy $P_\phi/L$:
$r=0$, which corresponds to the point-like graviton, and
\begin{equation}
r=\Bigl(\frac{4(n^2+3n)P_\phi}{NT_0L{\tilde L}}\Bigr)^{1/4}
\end{equation}
which corresponds to the dual giant graviton. When $N\rightarrow\infty$ 
\begin{equation}
r\rightarrow \Bigl(\frac{8P_\phi}{T_0L{\tilde L}}\Bigr)^{1/4}=
{\tilde L} \Bigl(\frac{P_\phi}{{\tilde N}}\Bigr)^{1/4}
\end{equation} 
in agreement with the result in \cite{GMT}.

\subsection{The macroscopical calculation}

Using the action (\ref{M5w}) we can easily describe the previous configurations
in terms of an M5-brane wrapped on the $\chi$-direction and carrying $P_\phi$ and
$P_\chi$ momentum charges. $P_\chi$ is switched on by dissolving $N$ gravitons 
with momentum on this direction in the worldvolume of the M5-brane.
We find
\begin{equation}
S=-4\pi^2 T_4 \int d\tau \Bigl[r\Bigl(\frac{r^4}{8}+\frac{N}{r^2}\Bigr)
\sqrt{1+\frac{r^2}{{\tilde L}^2}
-L^2{\dot{\phi}}^2}-\frac18 \frac{r^6}{{\tilde L}}\Bigr]\, ,
\end{equation}
and, in Hamiltonian formalism
\begin{equation}
\label{Hmacd}
H=\sqrt{1+\frac{r^2}{{\tilde L}^2}}\sqrt{\frac{P_\phi^2}{L^2}+16\pi^4 T_4^2 r^2
\Bigl(\frac{r^4}{8}+\frac{N}{r^2}\Bigr)^2}-\frac{\pi^2 T_4}{2}
\frac{r^6}{{\tilde L}}\, .
\end{equation}
Clearly the condition $P_\chi=0$ is met for $N=0$ gravitons dissolved in the
worldvolume, which amounts to setting to zero the $N$-term in the squared-root.
Comparing to (\ref{Hgiantd}) this further justifies the limit taken in that expression,
yielding to the microscopical potential (\ref{Hmicd}). In fact, when 
$N\rightarrow\infty$ ($\Leftrightarrow n\rightarrow \infty$) we find perfect agreement
between (\ref{Hmacd}) and the microscopical potential (\ref{Hmicd}).

\section{Conclusions}

We have shown that the giant graviton configurations in $AdS_m\times S^n$ 
backgrounds that
involve 5-spheres are described microscopically in terms of gravitational waves
expanding into ``fuzzy 5-spheres'' which are defined as $S^1$ bundles over fuzzy
$CP^2$. The explicit construction can be done due to the fact that the action used
to describe the system of coincident waves contains a special $U(1)$ isometric
direction that can be identified with the $U(1)$ fibre. 

In this description
the gravitons expand into a longitudinal M5-brane which has four manifest dimensions
and one wrapped on the $U(1)$ direction.
 This brane carries quadrupolar magnetic moment with respect to the 6-form potential of
the background in the $AdS_4\times S^7$ case, or quadrupolar electric moment in the
$AdS_7\times S^4$ case. In both cases it has a non-vanishing angular momentum along
the spherical part of the geometry, $P_\phi$.
The details of the construction show that there is as
well a non-vanishing momentum along the $U(1)$ direction, which has to be set to zero
to find the right point-like graviton and giant graviton configurations in the background. 
We have seen that
in that case not only the radii of the giant gravitons but also the Hamiltonian 
that they minimize
agree with the macroscopical results in \cite{GST,GMT} for large number of gravitons.

Gravitational waves
propagating both along the spherical part of the geometry (the $\phi$ direction) and
the compact $U(1)$ direction can be described macroscopically in terms of
a longitudinal M5-brane with velocity $\dot{\phi}$
wrapped on the isometric direction. The action associated to this brane
can be easily constructed by just
uplifting the action of the D4-brane to M-theory, while maintaining  
the eleventh direction compact.
The 
$\int C^{(1)}F\wedge F$ term in the CS part of the D4-brane action is then
uplifted to $\int k^{-2}k^{(1)} F\wedge F$, with $F$ now associated to wrapped M2-branes 
ending on the M5-brane. Therefore a momentum charge along the compact direction is
simply switched on by
taking $F$ with non-vanishing instanton number. The comparison between this
description and our microscopical description shows exact agreement for large
number of gravitons. Moreover, this comparison can be used  
to clarify the right way to set to zero the momentum
along the compact direction to finally obtain the correct giant graviton configurations.

Our action for M-theory waves, therefore, provides an explicit Matrix action which is
solved by some sort of non-commutative 5-sphere.  Moreover, although we have not checked the
supersymmetry properties of our configurations, the agreement with the macroscopical
description of \cite{GST,GMT} suggests that they should occur as BPS solutions 
preserving the same half of the supersymmetries as the point-like graviton \cite{GMT}.
To our knowledge this would be the first example of a physical matrix model, 
coming up as the action
for coincident M-theory gravitational waves,  admitting some fuzzy 5-sphere as a 
supersymmetry preserving solution. 

Let us stress that the ``fuzzy 5-sphere'' that we have constructed
is defined as an $S^1$ bundle
over the fuzzy $CP^2$, and is therefore different from previous fuzzy 5-spheres 
discussed in the literature \cite{Ram1,Ram2,S-J}.
In particular, our solution does not
show $SO(6)$ covariance, this invariance being broken down to $SU(3)\times U(1)$, whereas
this is the case for the fuzzy 5-sphere in \cite{Ram1,Ram2,S-J}.
The $SO(6)$ invariance might still be present in a non-manifest way, after all
the $SO(6)$ covariance of the classical 5-sphere is also not explicit when
it is described as an $S^1$-bundle over $CP^2$.
Another difference is that our solution approaches neatly the classical $S^5$ in the large
$N$ limit, where all the non-commutativity disappears. This is not the case for the
fuzzy 5-sphere in \cite{Ram1,Ram2,S-J}. Indeed, the right dependence of the radius of
the 5-sphere giant graviton with $P_\phi$, $R\sim P_\phi^{1/4}$, is only achieved
within the $S^1$ bundle over $CP^2$ description, the corresponding dependence of the
fuzzy 5-sphere of \cite{Ram1,Ram2,S-J} being given by $R\sim P_\phi^{1/5}$.
Another difference is that our ``fuzzy $S^5$'' inherits its symplectic
structure from the K\"ahler form of the fuzzy $CP^2$, whereas in the construction in 
\cite{Ram1,Ram2} the bundle structure corresponds to a $CP^3$ base and a
$CP^2$ fibre. Therefore, there are clear differences between the two constructions.

Non-supersymmetric longitudinal M5-branes with $CP^2\times S^1$ topology have 
been obtained as 
explicit solutions of Matrix theory in \cite{NR}. 
Our longitudinal M5-branes, although similar in the explicit construction,
have $S^5$ topology, once the necessary twist in the fibre is taken into
account. This twist should provide the global extension of the local residual
supersymmetry found in \cite{NR}, in terms of spinors charged under the gauge
potential whose field strength is the K\"ahler form (see \cite{DLP,pope}).

Longitudinal 5-branes with other topologies have also been shown to arise as
solutions to Matrix theory in \cite{CLT,Ho,K1,K2} (see also the fuzzy funnel
solution in \cite{CMT}).
In general, to find these solutions it is
necessary to include additional Chern-Simons terms or mass terms. Very recently \cite{N}
there has been some speculation on how the fuzzy 5-sphere of \cite{Ram1,Ram2} 
might appear as a solution to the pp-wave Matrix model of \cite{BMN}. It would be interesting
to elucidate the relation between the new Chern-Simons coupling conjectured in this reference
and the dielectric couplings constructed in this paper once the Penrose limit is 
taken. This would help clarifying if indeed the resulting Matrix action would allow
for transverse 5-brane solutions \cite{MSR}.

\subsubsection*{Acknowledgements}

We would like to thank J.M. Figueroa-O'Farrill and S. Ramgoolam for useful discussions.
The work of B.J. is done as part of the program ``Ramon y Cajal'' of the
M.E.C. (Spain). He was also partially supported by the M.E.C. under contract FIS
2004-06823 and by the Junta de Andaluc\'{\i}a group FQM 101.  
The work of Y.L. and D.R-G.
has been
partially supported by CICYT grant BFM2003-00313 (Spain). 
D.R-G. was supported in part by a F.P.U.
Fellowship from M.E.C. He would like to thank the String Theory group
at Queen Mary College (London) for its hospitality while part of this work was done.

%%%%%%%%%%%%%%%%%%%%%%%%%%%%%%%%%%%%%%%%%%%%%%%%%%%%%%%%%

%%%%%%%%%%%%%%%%%%%%%%%%%%%%%%%%%%%%%%%%%%%%%%%%%%%%%%%%%%
%\newpage
%\appendix 
%\renewcommand{\theequation}{\Alph{section}.\arabic{equation}}
%\sect{Blabla}

%%%%%%%%%%%%%%%%%%%%%%%%%%%%%%%%%%%%%%%%%%%%%%%%%%%%%%%%%%
%\newpage

%%%%%%%%%%%%%%%%%%%%%%%%%%%%%%%%%%%%%%%%%%%%%%%%%%%%%%%
%\end{multicols}           %end multicolumns
%%%%%%%%%%%%%%%%%%%%%%%%%%%%%%%%%%%%%%%%%%%%%%%%%%%%%%%
\end{document}